\documentclass[floatfix,aps,prd,twocolumn,showpacs,10pt]{revtex4}
\usepackage{epsfig}
\usepackage{epsf}
\usepackage{amsmath}
\usepackage{amssymb}

\begin{document}

\preprint{hep-ph/0606273}

\title{Isospin and a possible interpretation of the newly observed $X(1576)$}

\author{Feng-Kun Guo,$^{1,2,5}$}
\author{Peng-Nian Shen$^{2,1,3,4}$}
\affiliation{$^1$Institute of High Energy Physics, Chinese Academy
of Sciences,
P.O.Box 918(4), Beijing 100049, China \\
$^2$CCAST(World Lab.), P.O.Box 8730, Beijing 100080, China\\
$^3$Institute of Theoretical Physics, Chinese Academy of Sciences, P.O.Box 2735, China\\
$^4$Center of Theoretical Nuclear Physics, National Laboratory of
Heavy Ion Accelerator, Lanzhou 730000, China\\
$^5$Graduate University of Chinese Academy of Sciences, Beijing
100049, China}

\date{\today}

\begin{abstract}
Recently, the BES collaboration observed a broad resonant structure
$X(1576)$ with a large width being around 800 MeV and assigned its
$J^{PC}$ number to $1^{--}$. We show that the isospin of this
resonant structure should be assigned to 1. This state might be a
molecule state or a tetraquark state. We study the consequences of a
possible $K^*(892)$-${\bar \kappa}$ molecular interpretation. In
this scenario, the broad width can easily be understood. By using
the data of $B(J/\psi\to X\pi^0)\cdot B(X\to K^+K^-)$, the branching
ratios $B(J/\psi\to X\pi^0)\cdot B(X\to \pi^+\pi^-)$ and
$B(J/\psi\to X\pi^0)\cdot B(X\to K^+K^-\pi^+\pi^-)$ are further
estimated in this molecular state scenario. It is shown that the
$X\to \pi^+\pi^-$ decay mode should have a much larger branching
ratio than the $X\to K^+K^-$ decay mode has. As a consequence, this
resonant structure should also be seen in the $J/\psi\to
\pi^+\pi^-\pi^0$ and $J/\psi\to K^+K^-\pi^+\pi^-\pi^0$ processes,
especially in the former process. Carefully searching this resonant
structure in the $J/\psi\to \pi^+\pi^-\pi^0$ and $J/\psi\to
K^+K^-\pi^+\pi^-\pi^0$ decays should be important for understanding
the structure of $X(1567)$.
\end{abstract}


\maketitle

Recently, the BES collaboration analyzed the $J/\psi\to K^+K^-\pi^0$
decay, and found a broad resonant structure in the $K^+K^-$
invariant mass spectrum. The pole position of the resonant structure
is $1576^{+49+98}_{-55-91}-i409^{+11+32}_{-12-67}$ MeV, the $J^{PC}$
number is $1^{--}$, but its isospin has not been assigned yet
\cite{bes06}. They also claimed that this broad structure (refer to
$X(1576)$ in the rest of the text) cannot be explained as any known
mesons or their mixing states. Although the contribution from the
subthreshold $\rho$ may give a significant influence on the partial
wave analysis of the $J/\psi\to K{\bar K}\pi$ decay \cite{wz06}, the
inclusion of this $\rho$ state or even other mesons cannot remove
the resonant structure at $1576$ MeV and meanwhile will produce
large systematic errors \cite{bes06}.

An important character of $X(1576)$ is that the width of about 800
MeV is much larger than the width of any known vector mesons. The
PDG data \cite{pdg04} show that the largest width of vector meson,
for instance the width of $\rho(1450)$, is about $400$ MeV. In the
two-body decay process, a $J^P=1^{--}$ vector meson could decay
either into two $J^P=0^{-}$ mesons or into one $J^P=0^{-}$ and one
$J^P=1^{\pm}$ (or $J^P=2^{+}$) mesons. In these decays, the decay
width of the $1^{--}$ meson would not be very large due to the P
wave suppression in the former case and due to the phase space
suppression in the later case. Therefore, it is difficult to find a
proper place in the conventional $q{\bar q}$ meson spectrum for such
a particle. The $X(1576)$ state cannot be a glueball state because a
vector glueball consists of at least three gluons and the lattice
calculation showed that the mass of the vector glueball should be
about 3.8 GeV \cite{latG}. The large width of $X(1576)$ prohibits it
to be assigned as a hybrid. A flux tube model calculation showed the
total width of the favorable decay modes of a $1^{--}$ hybrid at 2
GeV is much smaller than 800 MeV \cite{hybr}. Besides, one argued
that the width of a $1^{--}$ hybrid decaying into $K{\bar K}$
vanishes \cite{hybr}. It means that $X(1576)$ found in the $K^+K^-$
invariant mass spectrum cannot be assigned as a vector hybrid. The
width of $X(1576)$ is consistent with the argument that the width of
a multi-quark state who is falling apart should be at least 500 MeV
\cite{cpr03}. Thus, the room left for $X(1576)$ is the tetraquark
state and the meson-meson molecular state. At least, such structures
should be the dominant components in $X(1576)$.

In this letter, we firstly show that the isospin of the broad
structure should be 1. Then, we study some consequences of a
possible molecule configuration of the broad structure $X(1576)$,
i.e. $K^*(892)$-${\bar \kappa}$ molecule. We calculate the ratio of
the widths $\Gamma(X\to\pi^+\pi^-):\Gamma(X\to K^+K^-):\Gamma(X\to
K^+K^-\pi^+\pi^-)$, and further estimate the branching ratios of the
$J/\psi\to\pi^+\pi^-\pi^0$ and $J/\psi\to K^+K^-\pi^+\pi^-\pi^0$
decays through intermediate state $X(1576)$.


In order to determine the isospin of $X(1576)$, BES Collaboration
mentioned that the $J/\psi\to K_SK^{\pm}\pi^{\mp}$ decay should be
studied \cite{bes06}. In fact, because the isospin of the $K^+K^-$
system can be either 0 or 1, the isospin of $X(1576)$ can be
figured out by examining whether the $J/\psi\to X\pi^0$ decay
favors the isospin symmetry. Namely, if the decay violates the
isospin symmetry, the isospin should be 0, otherwise it should be
1.

If $J/\psi\to X\pi^0$ decay violates the isospin symmetry,
$J/\psi\to X\eta$ decay must favor the isospin symmetry, and
$J/\psi\to X\pi^0$ decay should occur through $\pi^0$-$\eta$ mixing.
Following Dashen's theorem \cite{dash}, $\pi^0$-$\eta$ mixing should
be
\begin{equation}
t_{\pi\eta}=\langle\pi^0|{\cal H}|\eta\rangle=-0.003 ~\text{ GeV}^2.
\end{equation}
Then, the ratio of the coupling constants $g_{J/\psi X\eta}$ and
$g_{J/\psi X\pi^0}$ is
\begin{equation}
|\frac{g_{J/\psi X\pi^0}}{g_{J/\psi X\eta}}| =
|\frac{t_{\pi\eta}}{m_{\pi^0}^2-m_{\eta}^2}| = 0.01,
\end{equation}
and the ratio of the branching ratios $B(J/\psi\to X\pi^0)$ and
$B(J/\psi\to X\eta)$ is
\begin{equation}
R_{\pi^0/\eta}\equiv\frac{B(J/\psi\to X\pi^0)}{B(J/\psi\to X\eta)}
\approx |\frac{g_{J/\psi X\pi^0}}{g_{J/\psi X\eta}}|^2 = 1 \times
10^{-4}.
\end{equation}
On the other hand, BES Collaboration measured \cite{bes06}
\begin{equation}
\label{eq:bkk} B(J/\psi\to X\pi^0)B(X\to K^+K^-) =
(8.5\pm0.6^{+2.7}_{-3.6})\times10^{-4}.
\end{equation}
If the isospin of $X(1576)$ could be 0, one should have
\begin{eqnarray}
B(J/\psi\to X\eta)B(X\to K^+K^-) &\approx& \frac{1}{R_{\pi^0/\eta}} B(J/\psi\to X\pi^0) \nonumber\\
& & \times B(X\to K^+K^-) \nonumber\\
&>& 1.
\end{eqnarray}
Clearly, it cannot be true. Therefore, one can conclude that the
isospin of $X(1576)$ should be 1, and the observed structure should
be the $(I,I_3)=(1,0)$ state of the iso-triplet. Although the state
should be named as $\rho(1576)$ according to the nomenclature in PDG
\cite{pdg04}, we still use $X(1576)$ as its name in the rest of the
letter for consistency.


Assuming the observed broad structure is a vector isovector state,
we consider the possibility assigning it as a $K^*(892)$-${\bar
\kappa}$ molecule state. The evidences of the controversial $\kappa$
have been observed in the analysis of the $K\pi$ scattering phase
shifts \cite{lass}, the Dalitz Plot Analysis of the Decay $D^+\to
K^-\pi^+\pi^+$ \cite{e791}, and the BES data of the $J/\psi$ decays
\cite{wu03,bugg,bes05}. Despite of much controversy about the mass
and the width of $\kappa$, even the existence of $\kappa$, and of
the large experimental error on the mass of $X(1576)$, in this
molecular scenario, the large width of $\kappa$ could lead to a
large width of $X(1576)$. Now, we estimate the decay widths of some
possible two-body hadronic decays of $X(1576)$. The possible decay
channels considered in this report are $X(1576)^0\to
K^+K^-,~\pi^+\pi^-,~K^+K^-\pi^+\pi^-$. For these channels, the
involved coupling constants are the same, so we can obtain the
ratios of of the three branching fractions without information of
the coupling constants. In the estimation, we take
$m_{\kappa}\approx750$ MeV and $\Gamma_{\kappa}\approx550$ MeV which
are predicted in the theoretical calculations \cite{kappa}. Using
the phase convention $|\bar{
\kappa}^0\rangle=-|I,I_3\rangle=-|1/2,1/2\rangle$, the molecular
state with $(I,I_3)=(1,0)$ can be written as
\begin{equation}
|X^0\rangle = \frac{1}{\sqrt{2}}|(K^*(892)^+\kappa^- -
K^*(892)^0{\bar \kappa}^0)_X\rangle,
\end{equation}
where the subscript $X$ denotes the $X$ state coupled by
$K^*(892)$ and ${\bar \kappa}$.

\begin{figure}[htb]
\begin{center}\vspace*{0.cm}
{\epsfysize=3.2cm \epsffile{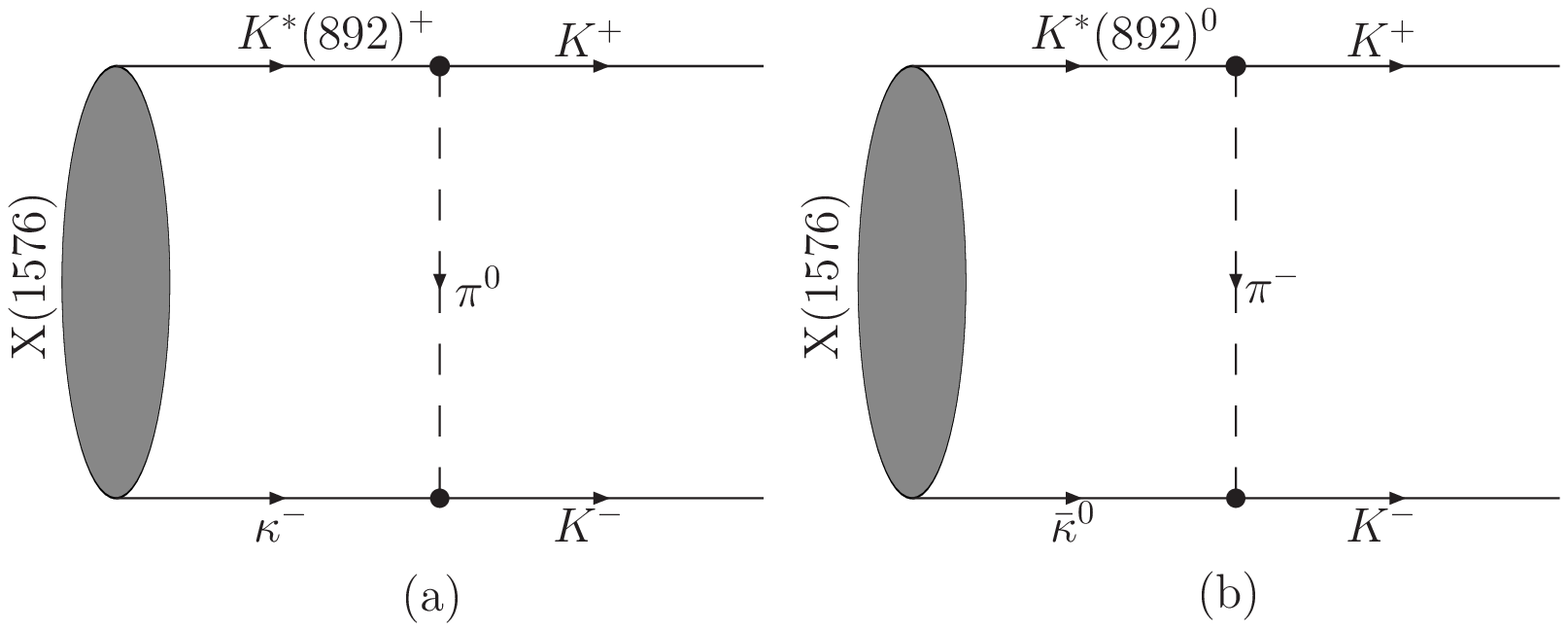}}%
\vglue 0cm\caption{\label{fig1}Decays of $X(1576)\to K^+K^-$. (a)
and (b) describe the decays of the $K^*(892)^+\kappa^-$ and
$K^*(892)^0{\bar \kappa}^0$ components of $X(1576)$,
respectively.}
\end{center}
\end{figure}

$X(1576)$ state can decay into $K^+K^-$ by exchanging non-strange
meson between the $K^*(892)$ and ${\bar \kappa}$. Constrained by the
parity and the angular momentum conservation laws, among light
mesons $\pi$, $\sigma$ and $\rho$, only $\pi$ can be exchanged. In
Figs. \ref{fig1}(a) and \ref{fig1}(b), we demonstrate the decays of
the $K^*(892)^+\kappa^-$ and $K^*(892)^0{\bar \kappa}^0$ components
of $X(1576)$ into $K^+K^-$, respectively.

The effective Lagrangian for $X(1576)$-$K^*(892)$-${\bar \kappa}$
coupling can be written as
\begin{equation}
{\cal L}_X = \frac{g_1}{\sqrt{2}} ({\bar \kappa}\vec{\tau}\cdot
\vec{X}^{\mu}K^{*}_{\mu} + \text{H.c.}),
\end{equation}
where the fields of isospin multiplets are:
\begin{equation}
{\bar \kappa} = (\kappa^-~{\bar \kappa}^0),
~%
K^*_{\mu} =
\begin{pmatrix} K^{*+}_{\mu} \\ K^{*-}_{\mu} \end{pmatrix},
~%
\vec{\tau}\cdot \vec{X}_{\mu} =
\begin{pmatrix}
X^0_{\mu} & \sqrt{2}X^-_{\mu} \\
\sqrt{2}X^+_{\mu} & -X^0_{\mu} \end{pmatrix}.
\end{equation}
The $K^*$-$K$-$\pi$ coupling can be obtained from the $SU(3)$
symmetric Lagrangian \cite{ld90}
\begin{equation}
{\cal L}_{VPP} = \frac{i}{2} G_V
\text{Tr}([P,\partial_{\mu}P]V^{\mu}),
\end{equation}
where $P$ and $V_{\mu}$ are $3\times3$ matrices,
$P=\sum_{a=1}^8\lambda^aP^a$ and
$V_{\mu}=\sum_{a=1}^8\lambda^aV_{\mu}^a$ with $\lambda^a$ being
the Gell-Mann matrices. And the $\kappa$-$K$-$\pi$ coupling can be
obtained from the effective Lagrangian \cite{bf99}
\begin{equation}
{\cal L}_{\kappa K\pi} = -\frac{1}{\sqrt{2}} g_{\kappa K\pi}
(\partial_{\mu}{\bar K}\vec{\tau}\cdot
\partial^{\mu}\vec{\pi} \kappa + \text{H.c.}).
\end{equation}
It should be mentioned that these Lagrangians were used and tested
by the others in the meson-meson scattering and the meson decay
processes \cite{ld90}. Because near the threshold of a particular
channel, the interaction between the particles in the channel is
rather weak, the t-channel contribution would dominate and the
weakly bound approximation would be a good approximation, namely
this loosely bound system would easily break up to the free
particles of which the corresponding component of the system was
composed. We also enforce a bound state condition to the system by
restricting the invariant mass in the decay channel being
equivalent to the mass of the decay state. Thus, near threshold a
meson-meson interaction model would be appropriate for a loosely
bound system.

With these effective Lagrangians and the isospin symmetry, one
finds
\begin{equation}
{\cal M}(X\to K^+K^-)_b = -2{\cal M}(X\to K^+K^-)_a,
\end{equation}
where the subscripts $a$ and $b$ denote Fig. \ref{fig1}(a) and
\ref{fig1}(b), respectively. Then, the total decay amplitude of
the $X(1576)\to K^+K^-$ process reads
\begin{equation}
{\cal M}_{KK}\equiv{\cal M}(X\to K^+K^-) = -{\cal M}(X\to K^+K^-)_a.
\end{equation}
In calculating the partial decay width of the $X(1576)\to K^+K^-$
process, a 3-meson loop, containing $K^*$, $\kappa$ and $\pi$
propagators, is involved in the transition amplitude (see Fig.
\ref{fig1}). In order to simplify the calculation, the on-shell
approximation for the $K^*$ propagator is employed in dealing with
loop integration, namely, the denominator of the $K^*$ propagator
is replaced by $(-i\pi)\delta(k^2-m_{K^*}^2)$. A similar
replacement for the $\kappa$ propagator is also performed. Then,
the decay amplitude can be re-written as
\begin{eqnarray}
{\cal M}_{KK} &=& \frac{\pi^2}{2} g_1g_{K^*K\pi}g_{\kappa K\pi}
\int\frac{d^4k}{(2\pi)^4} \delta(k^2-m_{K^*}^2) \nonumber\\
& & \delta((p-k)^2-m_{\kappa}^2) p_2\cdot(k-p_1) \nonumber\\
& & \varepsilon(X)_{\mu}
\frac{-g^{\mu\nu}+k^{\mu}k^{\nu}/m_{K^*}^2}{(k-p_1)^2-m_{\pi}^2}(2p_1-k)_{\nu},
\end{eqnarray}
where $g_{K^* K\pi}=G_V$, and $p,~k,~p_1~p_2$ are 4-momenta of the
$X(1576)$, $K^*(892)^+$, $K^+$ and $K^-$ particles, respectively.

\begin{figure}[htb]
\begin{center}\vspace*{0.cm}
{\epsfysize=3.2cm \epsffile{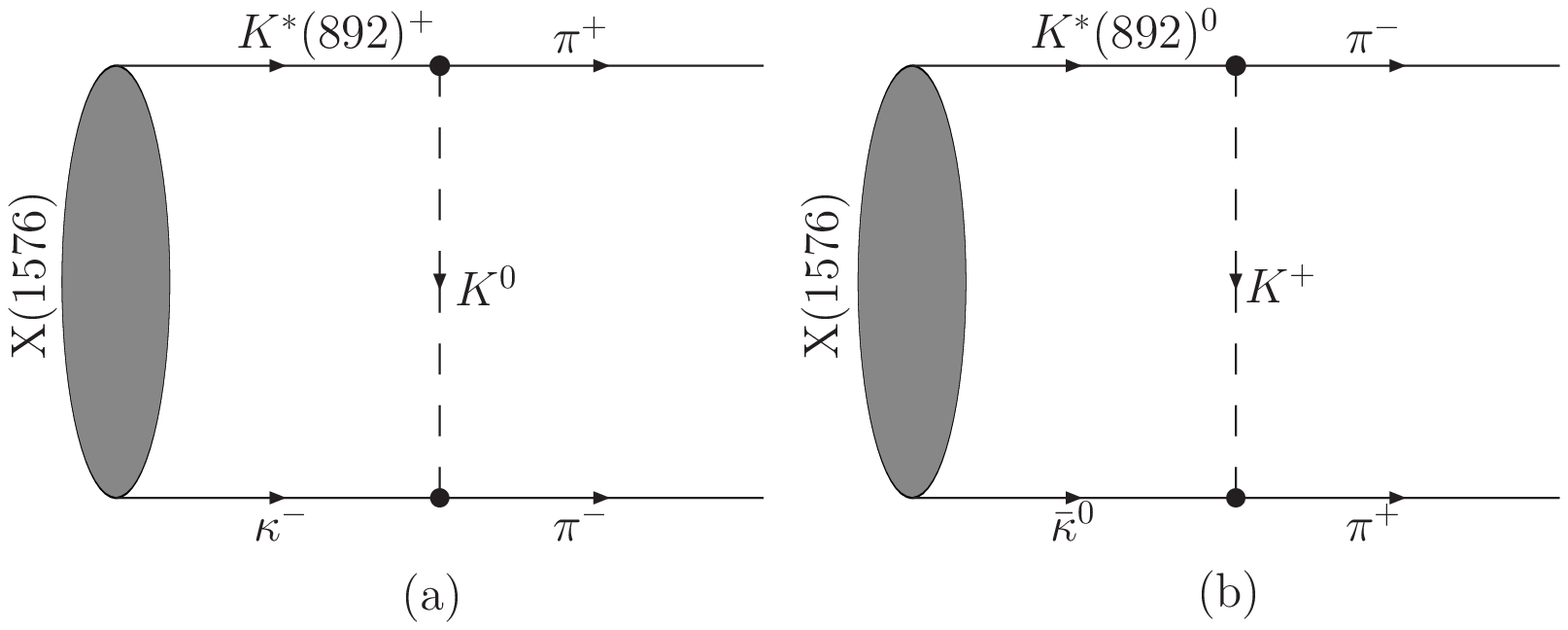}}%
\vglue 0cm\caption{\label{fig2}Decay of $X(1576)\to \pi^+\pi^-$
(a) and (b) denote the decays of the $K^*(892)^+\kappa^-$ and
$K^*(892)^0{\bar \kappa}^0$ components of $X(1576)$,
respectively.}
\end{center}
\end{figure}

For the $X(1576)\to\pi^+\pi^-$ process, $X(1576)$ can decay via
$K$ exchange shown in Fig. \ref{fig2}.
The decay amplitude for Fig. \ref{fig2}(a) can be obtained from
that for Fig. \ref{fig1}(a) by replacing $m_{\pi}$ with $m_K$
\begin{equation}
{\cal M}(X\to \pi^+\pi^-)_a = -2{\cal M}(X\to K^+K^-)_a
|_{m_{\pi}\leftrightarrow m_K},
\end{equation}
while the decay amplitude for Fig. \ref{fig2}(b) can be obtained
from ${\cal M}(X\to \pi^+\pi^-)_a$ by replacing the momentum of
$\pi^+$ with the momentum of $\pi^-$
\begin{equation}
{\cal M}(X\to \pi^+\pi^-)_b = -{\cal M}(X\to \pi^+\pi^-)_a
|_{p_{\pi^+}\leftrightarrow p_{\pi^-}}.
\end{equation}
Then, the total decay amplitude of the process
$X(1576)\to\pi^+\pi^-$ can be written as
\begin{equation}
{\cal M}(X\to \pi^+\pi^-) = {\cal M}(X\to \pi^+\pi^-)_a +{\cal
M}(X\to \pi^+\pi^-)_b.
\end{equation}

\begin{figure}[htb]
\begin{center}\vspace*{0.cm}
{\epsfysize=3.2cm \epsffile{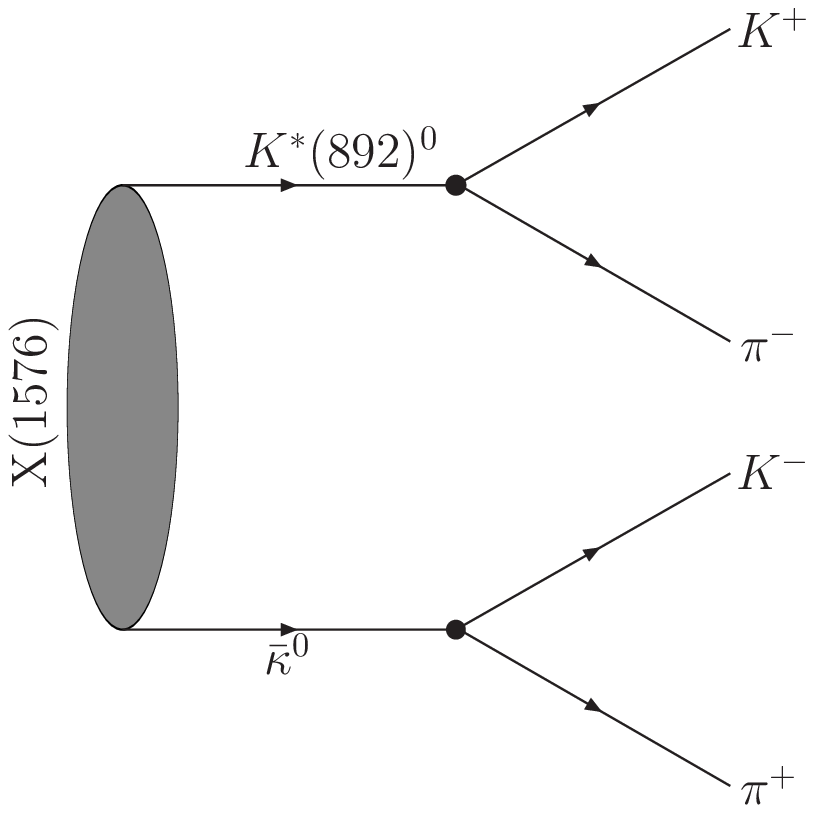}}%
\vglue 0cm\caption{\label{fig3}Decay of $X(1576)\to
K^+K^-\pi^+\pi^-$.}
\end{center}
\end{figure}

Since the most dominant decay channel for $K^*(892)$ and $\kappa$ is
the $K\pi$ channel, it is interesting to study the $X(1576)\to
KK\pi\pi$ decays. Because in the experiment, the charged particles
in the final state are much easier to be detected than the neutral
ones, we only calculate the $X(1576)\to K^+K^-\pi^+\pi^-$ decay. In
this decay, only the $K^*(892)^0$-${\bar \kappa}^0$ component
provides non-zero contribution. The corresponding decay diagram is
shown in Fig. \ref{fig3}.
The decay amplitude of this process reads
\begin{eqnarray}
{\cal M}_{KK\pi\pi} &\equiv& {\cal M}(X\to K^+K^-\pi^+\pi^-)
\nonumber\\
&=& -i g_1g_{K^*K\pi}g_{\kappa K\pi} \frac{p_3\cdot
p_4}{m_{34}^2-m_{\kappa}^2+im_{\kappa}\Gamma_{\kappa}(m_{34})}
\nonumber\\
& \times& \frac{\varepsilon(X)_{\mu}
(-g^{\mu\nu}+p_{12}^{\mu}p_{12}^{\nu}/m_{12}^2) (p_1-p_2)_{\nu}}
{m_{12}^2-m_{K^*}^2+im_{K^*}\Gamma_{K^*}(m_{12})},
\end{eqnarray}
where $p_1,~p_2,~p_3$ and $p_4$ represent the 4-momenta of $K^+$,
$\pi^-$, $K^-$ and $\pi^+$, respectively. $p_{12}=p_1+p_2$ is the
momentum of the $K^+\pi^-$ system, and $m_{12}$ and $m_{34}$ are
the invariant masses of the $K^+\pi^-$ and $K^-\pi^+$ systems,
respectively. The energy dependent widths of $K^*$ and $\kappa$
are
\begin{eqnarray}
\Gamma_{K^*}(m_{12}) &=& \Gamma_{K^*0}\frac{m_{K^*}^2|\vec{p}_1|^3}
{m_{12}^2|\vec{p}_{10}|^3}, \nonumber \\
\Gamma_{\kappa}(m_{34}) &=&
\Gamma_{\kappa0}\frac{m_{\kappa}^2|\vec{p}_3|}
{m_{34}^2|\vec{p}_{30}|} (\frac{m_{34}^2-m_K^2-m_{\pi}^2}
{m_{\kappa}^2-m_K^2-m_{\pi}^2})^2,
\end{eqnarray}
where $\Gamma_{K^*0}$ and $\Gamma_{\kappa0}$ are the decay widths
of $K^*(892)$ and $\kappa$ at rest, respectively, their values are
taken to be $\Gamma_{K^*0}=51$ MeV and $\Gamma_{\kappa0}=550$ MeV,
respectively,
\begin{eqnarray}
\label{eq:pcm} |\vec{p}_1| = \frac{1}{2m_{12}}
\sqrt{(m_{12}^2-(m_K+m_{\pi})^2)(m_{12}^2-(m_K-m_{\pi})^2)},
\nonumber\\
|\vec{p}_3| = \frac{1}{2m_{34}}
\sqrt{(m_{34}^2-(m_K+m_{\pi})^2)(m_{34}^2-(m_K-m_{\pi})^2)},\nonumber\\
\end{eqnarray}
and $|\vec{p}_{10}|$ and $|\vec{p}_{30}|$ can be obtained by
replacing $m_{12}$ and $m_{34}$ with $m_{K^*}$ and $m_{\kappa}$,
respectively.

Then the partial widths can be calculated by using the formula
\cite{pdg04}
\begin{equation}
\Gamma = \frac{1}{2m_X}\int \overline{\sum}|{\cal M}|^2(2\pi)^4
\delta(p-\sum_{i=1}^n p_i) \prod_{i=1}^n
\frac{d^3p_i}{(2\pi)^32E_i},
\end{equation}
where $\overline{\sum}$ denotes the average over the polarization
directions of the $X(1576)$ state.

Although the value of $g_1$ is not known, the ratio of the partial
widths of three decay modes $X(1576)\to\pi^+\pi^-$, $X(1576)\to
K^+K^-$ and $X(1576)\to K^+K^-\pi^+\pi^-$ can be evaluated as
\begin{eqnarray}
\label{eq:ratio} \Gamma(X\to\pi^+\pi^-) &\!\!\!:&\!\!\! \Gamma(X\to
K^+K^-) : \Gamma(X\to
K^+K^-\pi^+\pi^-) \nonumber\\
&\!\!\!\approx &\!\! 19:1:0.18.
\end{eqnarray}

Whether the estimated decay width ratio of the $\pi^+\pi^-$ decay
mode to the $K^+K^-$ decay mode is reliable can be checked by
comparing with the ratio in a naive estimation. In the later
estimation, if neglecting the mass difference between $\pi$ and $K$,
one can naively expect a ratio of 16, because comparing with ${\cal
M}(X\to K^+K^-)_a$, there is an extra factor of 2 in ${\cal M}(X\to
\pi^+\pi^-)_a$ due to the isospin symmetry, and Fig. \ref{fig2}(a)
and \ref{fig2}(b) give the same contribution.

Note that because the dominant decay mode of the $\kappa$ is $K\pi$
and the decay amplitude of $\kappa\to K\pi$ is proportional to  the
coupling constant $g_{\kappa K\pi}$, the large width of the $\kappa$
requires that $g_{\kappa K\pi}$ is large. The same constant
$g_{\kappa K\pi}$ appears in all of the above three partial widths,
and hence it will induce a large width of the $X(1576)$.

The above decay ratio also shows that in the $K^*(892)$-${\bar
\kappa}$ molecular scenario, comparing with the $X(1576)\to K^+
K^-$ decay, $X(1576)\to\pi^+\pi^-$ decay is much more favorable,
and the $X(1576)\to K^+K^-\pi^+\pi^-$ decay is suppressed by the
four-body phase space.

By using the above resultant ratio in Eq.(\ref{eq:ratio}) and the
measured branching ratio $B(J/\psi\to X\pi^0)B(X\to K^+K^-)$
\cite{bes06} in Eq. (\ref{eq:bkk}), the following branching ratios
can be estimated:
\begin{equation}
B(J/\psi\to X\pi^0)B(X\to \pi^+\pi^-) \approx (0.8\text{-}2.2)\%
\end{equation}
and
\begin{equation}
B(J/\psi\to X\pi^0)B(X\to K^+K^-\pi^+\pi^-) \approx
(0.8\text{-}2.1)\times10^{-4}.
\end{equation}
Comparing with the previously measured branching ratios
$B(J/\psi\to\pi^+\pi^-\pi^0)=(1.5\pm0.2)\%$ and $B(J/\psi\to
K^+K^-\pi^+\pi^-\pi^0)=(1.2\pm0.3)\%$ \cite{pdg04,mark2}, there is
still a room for interpreting the $X(1576)$ state as a
$K^*(892)$-${\bar \kappa}$ molecular state. If this state really
exists, the intermediate state $X(1576)$ should provide a dominant
contribution to the decay width of $J/\psi\to\pi^+\pi^-\pi^0$, but
not significant contribution to the decay width of $J/\psi\to
K^+K^-\pi^+\pi^-\pi^0$. Therefore, in the invariant mass spectrum
of $\pi^+\pi^-$, one should observe this resonant structure even
easier. Moreover, this feature can somehow be used to distinguish
the structure of the $X(1576)$ state. In the $qs{\bar q}{\bar s}$
($q=u,d$) tetraquark model, the $\pi\pi$ decay mode would be much
suppressed \cite{kl06,dy06}. In the $K{\bar K}$ molecular state
model, such decay modes would also be suppressed in comparison
with the $K{\bar K}$ decay mode since the non-strange decay modes
will take place through $K{\bar K}\to\pi\pi$ conversion.
Therefore, the relatively larger branching ratio of the
non-strange decay mode is a strong signal of the $K^*{\bar
\kappa}$ molecular model. We suggest to search the resonant
structure in the $J/\psi\to\pi^+\pi^-\pi^0$ decay channel by using
the high statistic BESII data. It is also valuable to re-analyze
the $K^+K^-\pi^+\pi^-$ invariant mass spectrum carefully by using
the data of the $J/\psi\to K^+K^-\pi^+\pi^-\pi^0$ decay, although
it is not so easy.

The $X$ state can also decay into other channels. For instance,
$\phi\pi$ channel would not be forbidden by any symmetry and can
take place through quark recombination, namely the $s$ quark in
$K^*$ and the ${\bar s}$ anti-quark in ${\bar \kappa}$ can be
combined into a $\phi$ meson, and the rest quark and anti-quark
can be combined into a $\pi$ meson. The decay rate of such a
channel in this scenario should further be investigated.

In summary, by utilizing the $\pi^0$-$\eta$ mixing mechanism of the
isospin violated decay, we can assign the isospin of the newly
observed broad structure $X(1576)$ to 1 definitely. If the observed
broad structure is a physical state, it is difficult to interpret it
as a conventional $q{\bar q}$ vector meson or a glueball or a hybrid
state. We propose a possible $K^*(892)$-${\bar \kappa}$ molecular
state interpretation. In this scenario, the large width can be
understood easily. We further estimate the decay properties of
$X(1576)$. Our results show that the $X(1576)\to\pi^+\pi^-$ decay
mode is much more favorable than the $X(1576)\to K^+K^-$ decay mode.
By comparing with the $J/\psi$ decay data taken previously, this
resonant structure should also appear in the
$J/\psi\to\pi^+\pi^-\pi^0$ and $J/\psi\to K^+K^-\pi^+\pi^-\pi^0$
decays in this scenario. In order to confirm this state, we suggest
to re-analyze the data of the $J/\psi\to\pi^+\pi^-\pi^0$ and
$J/\psi\to K^+K^-\pi^+\pi^-\pi^0$ decays collected at BESII.
However, there are other possible molecule and tetraquark
configurations \cite{kl06,dy06} that will result in different
predictions. The concrete consequences of other decay channels and
other possible molecule configurations, such as $K^+K^-$, will be
studied in the future.

\begin{acknowledgments}
We thank S. Jin and C.-P. Shen for discussion on experimental
status, and thank B.-S. Zou for useful discussion.  This work is
partially supported by the NSFC grant Nos. 10475089, 10435080, CAS
Knowledge Innovation Key-Project grant No. KJCX2SWN02 and Key
Knowledge Innovation Project of IHEP, CAS (U529).
\end{acknowledgments}


\begin{thebibliography}{99}
\bibitem{bes06}  M. Ablikim, {\it et al.} (BES Collaboration),
                hep-ex/0606047, submitted to Phys. Rev. Lett..

\bibitem{wz06}  F.Q. Wu and B.S. Zou,
                Phys. Rev. D {\bf 73}, 114008 (2006).

\bibitem{pdg04} S. Eidelman {\it et al.} (Particle Data Group),
                Phys. Lett. B {\bf 592}, 1 (2004).

\bibitem{latG}  C.J. Morningstar and M. Peardon,
                Phys. Rev. D {\bf 60}, 034509 (1999);
                Y. Chen et al.,
                Phys. Rev. D {\bf 73}, 014516 (2006).

\bibitem{hybr}  N. Isgur, R. Kokoski, and J. Paton,
                Phys. Rev. Lett. {\bf 54}, 869 (1985);
                P.R. Page, E.S. Swanson, and A.P. Szczepaniak,
                Phys. Rev. D {\bf 59}, 034016 (1999).

\bibitem{cpr03} S. Capstick, P.R. Page, and W. Roberts,
                Phys. Lett. B {\bf 570}, 185 (2003).

\bibitem{dash}  R. Dashen,
                Phys. Rev. {\bf 183}, 1245 (1969).

\bibitem{lass}  D. Aston {\it et al.},
                Nucl. Phys. {\bf B296},253 (1988).

\bibitem{e791}  E791 Collaboration, E. M. Aitala {\it et al.},
                Phys. Rev. Lett. {\bf 89}, 121801 (2002).

\bibitem{wu03}  N. Wu, International Symposium on Hadron Spectroscopy,
                Chiral Symmetry and Relativistic Description of Bound Systems,
                Tokyo, Japan, February 24-26, 2003.

\bibitem{bugg}  D. V. Bugg, Phys. Rept. {\bf 397}, 257 (2004).

\bibitem{bes05} BES Collaboration, M. Ablikim {\it et al.},
                Phys. Lett. B {\bf 633}, 681 (2006).

\bibitem{kappa} E.g., E. van Beveren {\it et al.},
                Z. Phys. C {\bf 30}, 615 (1986);
                D.V. Bugg,
                Phys. Lett. B {\bf 632}, 471 (2006);
                F.-K. Guo {\it et al.},
                Nucl. Phys. {\bf A773}, 78 (2006);
                Z.Y. Zhou and H.Q. Zheng,
                hep-ph/0603062.

\bibitem{ld90}  D. Lohse, J.W. Durso, K. Holinde, and J. Speth,
                Nucl. Phys. {\bf A516}, 513 (1990).

\bibitem{bf99}  D. Black, A.H. Fariborz, F. Sannino, and J. Schechter,
                Phys. Rev. D {\bf 59}, 074026 (1999).

\bibitem{mark2} M.B.E. Franklin {\it et al.},
                Phys. Rev. Lett. {\bf 51}, 963 (1983).

\bibitem{kl06}  M. Karliner and H.J. Lipkin,
                hep-ph/0607093.

\bibitem{dy06}  G.-J. Ding and M.-L. Yan,
                hep-ph/0607253.

\end{thebibliography}
\end{document}